\documentclass{article}

\usepackage{arxiv}

\usepackage[utf8]{inputenc} 
\usepackage[T1]{fontenc}    
\usepackage{hyperref}       
\usepackage{url}            
\usepackage{booktabs}       
\usepackage{amsfonts}       
\usepackage{nicefrac}       
\usepackage{microtype}      
\usepackage{lipsum}		
\usepackage{graphicx}
\usepackage{natbib}
\usepackage{doi}

\title{Data Augmentation Strategies for Improving Sequential Recommender Systems}

\date{September 9, 2021}	
\date{} 					

\author{{\hspace{1mm}Joo-yeong Song}\\
	GSCST\\
	Seoul National University\\
	Republic of Korea\\
	\texttt{moitie@snu.ac.kr} \\
	\And
	{\hspace{1mm}Bongwon Suh}\thanks{Corresponding Author}\\
	GSCST\\
	Seoul National University\\
	Republic of Korea\\
	\texttt{bongwon@snu.ac.kr} \\
}

\hypersetup{
pdftitle={Data Augmentation Strategies for Improving Sequential Recommender Systems},
pdfsubject={cs.LG, cs.IR},
pdfauthor={Jooyeong Song, Bongwon Suh},
pdfkeywords={Information systems, Recommender systems, Data augmentation},
}

\begin{document}
\maketitle

\begin{abstract}
  Sequential recommender systems have recently achieved significant performance improvements with the exploitation of deep learning (DL) based methods. However, although various DL-based methods have been introduced, most of them only focus on the transformations of network structure, neglecting the importance of other influential factors including data augmentation. Obviously, DL-based models require a large amount of training data in order to estimate parameters well and achieve high performances, which leads to the early efforts to increase the training data through data augmentation in computer vision and speech domains. In this paper, we seek to figure out that various data augmentation strategies can improve the performance of sequential recommender systems, especially when the training dataset is not large enough. To this end, we propose a simple set of data augmentation strategies, all of which transform original item sequences in the way of direct corruption and describe how data augmentation changes the performance. Extensive experiments on the latest DL-based model show that applying data augmentation can help the model generalize better, and it can be significantly effective to boost model performances especially when the amount of training data is small. Furthermore, it is shown that our proposed strategies can improve performances to a better or competitive level to existing strategies suggested in the prior works.
\end{abstract}


\section{Introduction}
Sequential recommendation has now been more widely studied, characterized by its well-consistency with real-world recommendation situations. Recently, deep learning (DL) based methods, which have contributed to significant performance improvement in diverse fields, have been adopted in sequential recommendation. Sequential recommender systems basically take sequences of items as their input, which is similar to natural language processing(NLP) that generally regards word sequences as its input unit. Based on this structural similarity, DL architectures such as recurrent neural networks (RNN) and attention mechanisms are extensively applied to sequential recommendation, delivering notable performance. Also, various methods based on other DL-architectures such as convolutional neural networks (CNN) and variational auto-encoder (VAE) have been increasingly adopted, demonstrating state-of-the-art performance.

Although it is clear that various DL-based methods have been actively introduced in sequential recommendation, most of them only focus on the transformations of model structure. However, according to previous works\citep{Fang20}, there are multiple factors influencing the performance of DL-based models, including data augmentation in the data processing step, and choice of loss function as a model training strategy. In particular, DL-based models with high complexity generally require a large amount of training data in order to estimate model parameters effectively and achieve high performance. For this reason, in some domains including computer vision (CV), there have been considerable early efforts to increase the size of training dataset through augmentation, which is currently established as a standard preprocessing technique for accomplishing high performance\citep{Krizhevsky12}. Yet, it is not sufficiently verified if and how data augmentation is effective in improving the performance of recommender systems.

In this paper, we aim to show that various data augmentation strategies can improve the performance of sequential recommender systems, especially with a limited amount of training dataset. To this end, we propose a simple set of data augmentation strategies including 1)Noise Injection, 2)Redundancy Injection, 3)Item Masking, 4)Synonym Replacement, all of which transform original item sequences with direct manipulation, and describe how data augmentation changes the performance through our extensive experiments based on the state-of-the-art recommendation model. The experiments demonstrate that overall performance improvements are achieved by the application of our proposed data augmentation strategies. It is notable that the performance improvement can be large if the size of the dataset is relatively small. This result suggests that our strategies can be particularly effective to boost the performance, especially for the cold-start situations in sequential recommender systems which do not have sufficient dataset in the early stage. 

The contributions of this study can be summarized as follows:
\begin{itemize}
\item We verify through quantitative experiments that data augmentation technology can improve DL-based sequential recommendation performances. We also demonstrate that it can be significantly effective to boost model performances especially when the amount of training data is small.
\item We describe how the performance varies by the extensive applications of various data augmentation strategies. In particular, we propose a simple set of augmentation strategies that directly manipulate original sequences, in addition to the subset-selection based ones suggested in prior works. It is shown that our suggested strategies can improve the performance to a better or competitive level with existing strategies.
\item We suggest the possibility of further performance improvements of other current state-of-the-art models, where data augmentation is applied in the preprocessing step while maintaining the overall model architecture. In this regard, we expect that future works can verify if data augmentation can serve as an universal preprocessing technique in the design of recommender systems. 
\end{itemize}

\section{Related Works}
\subsection{Sequential Recommendation}

Traditional recommendation aims to model general and global preferences of users based on the assumption that user-item relationships are not dynamic but static, which is intrinsically bound to ignore the order of user-item interactions in the model. Accordingly, other special approaches have been required for sequential recommendations, mainly concentrating on finding sequential patterns from data in the consideration of the item order\citep{Wang19survey}.

The earliest approach on sequential recommendation is Markov Chain (MC), which assumes that the next action is conditioned on only the few previous actions. MC aims to learn the transition patterns between items based on an item-item transition matrix applied equally to all users. {\itshape FPMC}\citep{Rendle10} is marked by its extensive approach that combines a 1st-order MC model with general matrix factorization, building a user-item-item tensor and factorizing it into a user-item matrix and a personalized transition matrix. The successively proposed {\itshape Fossil}\citep{He16} is an improved version of {\itshape FPMC} by extending factored MCs into higher orders. 

Recently, researchers have found that DL-based methods can be effective for sequential recommendation. RNN-based models are the most popular due to their structural commonality in which they both generally take their input in the form of sequences and model them step-by-step. The most pioneering DL-based method is {\itshape GRU4Rec}\citep{Hidasi15}, which utilize gated recurrent units (GRU) to model sequences. After that, an updated model {\itshape GRU4Rec+}\citep{Hidasi18} is suggested by adopting a different loss function and a sampling strategy. 
Other advanced approaches originally suggested in NLP, such as attention mechanisms, have also been increasingly brought and exploited in sequential recommendation. For example, {\itshape Transformer}\citep{Vaswani17} and {\itshape BERT}\citep{Devlin18}, purely composed of self-attention modules, are the most outstanding models in NLP, receiving attention for not only achieving state-of-the-art performance in multiple NLP tasks but also changing the landscape of the research field. {\itshape SASRec}\citep{Kang18} is the first method that seeks to apply the approaches of {\itshape Transformer} to sequential recommendation tasks. The model focuses on adopting the encoder structure of {\itshape Transformer} based of self-attention mechanisms in order to learn the latent representations of items in the sequences and utilize them as the basis of item similarity computation. It is proved that {\itshape SASRec} can outperform other previous DL-based methods on several benchmark datasets and thus has been established as a representative baseline model in sequential recommendation as multiple recently suggested methods - such as {\itshape BERT4Rec}\citep{Sun19}, {\itshape TiSASRec}\citep{Li20}, {\itshape FISSA}\citep{Lin20}, {\itshape SSE-PT}\citep{Wu20} - are following its approaches large and small.

\citep{Meng20} is clearly differentiated from other works in that it focuses on data splitting strategies in the preprocessing step, not the transformations of network structures or modules. However, it is mainly interested rather in the impact of data splitting strategies on the evaluation of recommendation models than the improvement of performance. As such, it is evident that most of the existing works making use of DL-based approaches seek to improve the performance solely by designing better network architectures, without adequately considering other aspects of the approaches.

\subsection{Data Augmentation}

DL-based models require large datasets to properly estimate numerous parameters because otherwise they can produce much worse performance than other simpler approaches. As it is common, in practice, that the amount of data in hand is limited and the cost of acquiring additional data is high, one way to deal with this is data augmentation, a technique to create fake data based on the dataset currently available and add them to the training dataset\citep{Goodfellow16}.

Data augmentation is known to be particularly effective for classification problems. For object recognition tasks in the CV domain, some augmentation strategies have already proved to greatly improve generalization, thereby becoming a standard technique in the data preprocessing step. Since there should be more appropriate or inappropriate ways of augmenting by the characteristics of datasets, most researchers have paid attention to finding the best augmentation strategies for each specific benchmark dataset\citep{Shorten19}; for example, natural image datasets such as ImageNet, random cropping, image mirroring, and coloring shifting\citep{Krizhevsky12}. Also for speech recognition tasks, several effective augmentation strategies have been proposed to operate on the raw signals or log mel spectrogram of input audio, such as speed perturbation or noise injection\citep{Ko15,Park19}.
 
In NLP, progress in data augmentation is relatively slow due to the challenge of establishing generalized rules for language transformation. Yet, as the technique has recently been proved promising to boost performance in text classification tasks, interests are growing on the augmentation of natural language data. {\itshape EDA}\citep{Wei19} has presented a set of universal data augmentation strategies (i.e., synonym replacement, random insertion, random swap, and random deletion) and confirmed that the technique can lead to substantial improvement by randomly operating one of the strategies, being particularly helpful for smaller datasets. This study, which provides the methodological basis for the evaluation of the technique’s effectiveness, uses a similar method to ours, assuming low-resource situations by using only a restricted fraction of the original training dataset.
 
Even though data augmentation has been established in many domains as a standard technique to train more robust models, it is hard to say that it has been fully explored in recommendation. One pioneering work is  {\itshape AugCF}\citep{Wang19}, which utilizes data augmentation to alleviate the inefficiency of existing hybrid collaborative filtering methods. The key idea is using side information in the process of data augmentation, not using it directly as a simple additional input.

In sequential recommendation, there have also been only few works that have explored data augmentation\citep{Grbovic15,Wolbitsch19}. The earliest work is {\itshape Improved RNN}\citep{Tan16}, which proposed a two-step method to augment click sequences particularly on {\itshape YooChoose} dataset. The first step is treating all prefixes of the original sequence as new training sequences and next applying item dropout randomly to the augmented subsequences. Thanks to data augmentation, the extended model can enhance the performance of basic RNN models, thus becoming a standard preprocessing technique for the subsequent models that evaluate on the dataset. Besides, {\itshape CASER}\citep{Tang18} is a CNN-based model which applies horizontal and vertical convolutional filters to capture short-term preferences in item sequences. It operates the strategy of sliding window in order to apply CNN architecture to sequential data, splitting an original sequence of length  {\itshape m} into several subsequences with the same length (i.e. window) of smaller {\itshape L}, and generates $(m-L+1)$ subsequences from a sequence, resulting in an enlarged size of training dataset. In all of these previous works, augmented subsequences leave the order of original sequence intact. This makes their strategies dependent on the length of original sequence while our version is capable of augmenting data at any desirable size.

\section{Proposed Method}

In this paper, we propose a set of data augmentation strategies for sequential recommendation. To the best of our knowledge, we are the first study to comprehensively explore the effectiveness of augmenting data in sequential recommendation with extensive experiments. Note that we only focus on the data augmentation operated in the preprocessing step, with the other network architecture intact. In our experiments, we use {\itshape SASRec}\citep{Kang18}, a state-of-the-art model which applies self-attention mechanisms as the baseline model. Thus, the overall network architecture of the recommender system follows the original design of {\itshape SASRec} model.

\subsection{Problem Formulation}

Before the illustration of our suggested strategies, we first formulate the sequential recommendation problem as follows. Without loss of generality, we have a recommendation system with implicit feedback given by a set of users $U$ to a set of items $I$. For sequential recommendation, we denote the records of each user $u \in U$ as an item sequence (in the order of interaction time) as $S^u =\{ s^u_1,s^u_2,...,s^u_{|S^u|} \}$, $s^u \in I$. Our goal is to provide a recommendation list for each user $u$, in which we expect the next real interacted item $s^u_{|S^u|+1} \in I\verb|\|S^u$ to appear and be ranked as high as possible, hopefully the first.

\begin{table}
  \centering
  \caption{Classification of data augmentation strategies 
  for sequential recommendation.}
  \label{tab:clf}
  \begin{tabular}{ccc}
    \toprule
    Category & Strategy & Notes\\
    \midrule
    Direct Manipulation & Noise Injection (NI) & Suggested in this paper\\
     & Redundancy Injection (RI) & \\
     & Item Masking (IM) & \\
     & Synonym Replacement (SR) & \\
   \midrule
   Subset Selection & Subset Split (SS) & Suggested in prior works\\
    & Sliding Window (SW) & \\
   \bottomrule
\end{tabular}
\end{table}

\subsection{Augmentation Strategies}

In this section, we present the details of four augmentation strategies that can be operated for each given item sequence of a training dataset.

\begin{itemize}
\item Noise Injection (NI): Randomly choose an item (i.e., negative sample) not included in the original item sequence. Inject the negative sample into a random position in the sequence.
\item Redundancy Injection (RI): Randomly choose an item (i.e., positive sample) from the original item sequence. Inject the positive sample into a random position in the sentence.
\item Item Masking (IM): Randomly choose {\itshape k} items from the original item sequence. Mask the {\itshape k} items to exclude them from the training. The size of {\itshape k} is decided based on the sequence length {\itshape m} with the formula $k = p*m$ where {\itshape p} is a parameter indicating the percentage of the items to be masked.
\item Synonym Replacement (SR): Randomly choose an item from the original item sequence. Replace it with the most similar {\itshape s} items (i.e., synonyms).
\end{itemize}

To apply the NI or RI strategy, an item should be removed from the original sequence since the length of augmented item sequences should remain the same as the length of the original sequence due to the model structure. According to previous research, the latest item has the most significant impact on the prediction of the next item while the far-away items have a relatively low impact.\citep{Tang18,Wang19} Based on the lessons, our design removes the first item of the original sequence from the training, assuming that it would have the lowest impact on the next-item prediction.

Besides, to apply the SR strategy, it is necessary to set the criteria on item similarities to determine the most similar ones (i.e., synonyms) to each item. Thus, we decide to train the item embeddings first to compute the similarities. In this paper, we use the {\itshape Word2Vec}\citep{Mikolov13} algorithm prevalently utilized to acquire the pretrained word embeddings in NLP thanks to its easiness of implementation and computational efficiency. However, the algorithm has its intrinsic limitations where we cannot verify how accurately the similarities given by the trained item embeddings reflect the 'real' similarities between items. Therefore, our design generates augmented samples based on multiple similar ones for an item to be replaced, as the simple replacement with 'the most similar' one would have a higher risk of degrading the quality of the augmented data.

Note that for all augmentation strategies suggested here, original item sequences remain intact and ($N_{aug}-1$) artificial data samples are generated additionally, where {\itshape $N_{aug}$} is a parameter that indicates the number of augmented sequences per each original one. For example, if it is set to be $N_{aug} = 10$, the NI or RI strategy can generate 9 unique artificial sequences for an original item sequence and each will have only one-item-difference from the original sequence. In contrast, IM will generate 9 artificial sequences which have {\itshape k} different items from the original one, respectively. In case of the SR strategy, assuming to set $N_{aug} = 10$ and $s = 3$, it will randomly choose 3 items in the original sequence and replace each with its most similar 3 items, resulting in 9 different sequences having one-item-difference from the original one.

\section{Experiments}

In this section, we present our experimental settings and results to answer the following research questions: 
\begin{itemize}
\item RQ1: Can our proposed data augmentation strategies improve the performance of state-of-the-art baselines for sequential recommendation tasks?
\item RQ2: How do the different data augmentation strategies affect model performance differently?
\item RQ3: How do the size of data augmentation strategies affect model performance differently?
\end{itemize}

Note that we evaluate our methods through extensive comparison with two baseline situations: 1) No data augmentation is applied, 2) Data augmentation strategies suggested in prior works\citep{Tan16,Tang18} are applied.

\subsection{Baseline Models and Strategies}

As we seek to improve performance solely through the data augmentation operated in the preprocessing step, we choose to use a state-of-the-art model {\itshape SASRec}\citep{Kang18} for sequential recommendation tasks as our baseline. Also, since the model leverages only the item order information which is essential for general sequential recommendation models, it is adequate to verify if data augmentation can be a standard for various sequential recommendation models.

Besides, the compared augmentation methods are Subset Split (SR)\citep{Tan16} and Sliding Window (SW)\citep{Tang18}, both of which generate augmented data samples by using the subsets of the original sequences. SR implements data augmentation by splitting the original sequence of length {\itshape m} into shorter subsequences of length 1 \verb|~| {\itshape m}. In this experiment, we consider the number of $N_{aug}$ as the maximum size of augmentation if the original sequence length {\itshape m} is larger than $N_{aug}$ (i.e., $(m > N_{aug})$) in order to control the size of augmentation and apply it on the same scale for all training sequences as much as possible. Meanwhile, SW splits the original sequence of length {\itshape m} into subsequences with the same length {\itshape L} by sliding with a window of length {\itshape L}. Likewise, we set the parameter $N_{aug}$ fixed and adjust the window size {\itshape L} in accordance with the formula $N_{aug} = m - L + 1$.

\begin{table}
  \centering
  \caption{Basic Dataset Statistics.}
  \label{tab:dataset}
  \begin{tabular}{cccccc}
    \toprule
    Dataset&\#\ users&\#\ items&\#\ interactions&Avg. Length&Sparsity\\
    \midrule
    MovieLens-1M&6,040&3,416&999,611&165.50&95.16\\
    Amazon Games&64,073&33,614&568,508&6.88&99.97\\
    Gowalla&85,034&308,957&6,442,892&52.83&99.98\\
   \bottomrule
\end{tabular}
\end{table}

\subsection{Datasets}

We evaluate our methods on three benchmark datasets from real-world platforms which have different characteristics in terms of domain, size, sparsity, and average sequence length: MovieLens-1M, Amazon Games, and Gowalla. Summary statistics are shown in Table 2 above. Furthermore, we hypothesize that data augmentation can be more helpful for smaller datasets, so we delegate the restricted fraction of datasets by selecting a random subset of the full training set.

\begin{itemize}
\item MovieLens\citep{harper2015movielens}: A widely-used benchmark dataset especially for evaluating collaborative filtering algorithms. We use the version that includes 1 million user ratings (i.e., MovieLens-1M). It is known as a relatively dense dataset for its low sparsity and long sequence length in average.
\item Amazon Games\citep{He16,McAuley15}: Amazon is a series of datasets comprising large volume of product reviews crawled from {\itshape Amazon.com}. Among various categories, we use ‘Video Games’ with high sparsity and variability.
\item Gowalla\citep{cho2011}: A location-based social networking website where users share their locations by checking-in, labeled with timestamps. It has high sparsity but medium average sequence length.
\end{itemize}

\subsection{Experimental Setup}

To preprocess all datasets, we follow the procedure from \citep{He2017,Kang18,Rendle10} as follows: 1) we regard the presence of a review, rating, or check-in record as implicit feedback and use timestamps to determine the sequence order of actions; 2) we discard cold-start users and items with fewer than 5 actions; 3) we adopt the leave-one-out evaluation by splitting each dataset into three parts, i.e., using the most recent item of the sequence for testing, the second recent item for validation and the remaining for training. Also, other parameters which do not have a direct impact on data augmentation are set to follow the baseline\citep{Kang18}. Source codes for both preprocessing and implementation are available at \url{https://github.com/saladsong}

According to prior works\citep{Wei19}, as the size of augmentation $N_{aug}$ is an important factor that affects performance significantly, it is necessary to tune it adequately for each dataset. Basically, we set the default size as $N_{aug} = 10$ for all datasets and strategies and next conduct additional experiments with the size of $N_{aug} \in \{ 2,5,15 \}$ in order to assess the contribution of the parameter.

For the evaluation of performance, we adopt two common metrics: hit ratio (HR@10) and normalized discounted cumulative gain (NDCG@10). HR@10 refers to the ratio of the ground-truth items presented in the top 10 recommendation lists, while NDCG@10 considers the position and assigns higher weights to higher positions. To reduce computation, referencing \citep{Kang18,Li20}, we randomly sample 100 negative items for each user { u} and rank these items with the ground-truth item, thereby calculating HR@10 and NDCG@10 based on the rankings of these 101 items. For all experiments, we average the results from five different random seeds and determine the best performance based on the metric of NDCG@10.

\subsection{Training Set Sizing}

In general, there is more tendency of overfitting in case of smaller training datasets, which means regularization such as data augmentation could bring larger gains on smaller datasets. Following \citep{Wei19}, we conduct experiments using a restricted fraction of the available training data for all three datasets. We run training both with and without any augmentation for the following training set fractions (\verb|%|): \{10, 20, 30, 40, 50, 100 (full) \}. Each sub-dataset is generated by random sampling from the original full dataset. Thus, only the size of the dataset is scaled down while the other characteristics such as sparsity or average sequence length remain almost the same.

\subsection{Results}

{\itshape 4.5.1. Performance Comparison (RQ1).} The experiments show that data augmentation on all datasets can lead to the improvement of sequential recommendation performance. The summary of the results is shown in  Table~\ref{tab:res_main}, illustrating the best strategy and its results for each sub-dataset with the augmentation size $N_{aug} = 10$. Note that the best performance for each case is counted based on NDCG@10. Although the performance varies depending on the augmentation strategies applied, injection strategies and Sliding Window (SW) have been found to be most effective overall. 

\begin{table*}[h]
  \centering
  \caption{Recommendation Performances with the Best Augmentation Strategies (NDCG@10)}
  \label{tab:res_main}
  \begin{tabular}{c|ccc|ccc|ccc}
    \toprule
    Size & \multicolumn{3}{c|}{MovieLens-1M} & \multicolumn{3}{c|}{Amazon Games} & \multicolumn{3}{c}{Gowalla} \\
    \cline{2-10}
      & Base&Prior(SW)&Ours(RI)& Base&Prior(SW)&Ours(NI)& Base&Prior(SW)&Ours(NI) \\
    \midrule
    10\%&0.3114&0.3936&\textbf{0.4003}&0.1830&0.1991&\textbf{0.2350}&0.4981&0.5060&\textbf{0.5149}\\
     & &(+26.4\%)&\textbf{(+28.5\%)}& &(+8.8\%)&\textbf{(+28.4\%)}& &(+1.6\%)&\textbf{(+3.4\%)}\\
     
    20\%&0.4305&0.4576&\textbf{0.4609}&  0.2462&0.2965&\textbf{0.3240}&0.5612&0.5718&\textbf{0.5821}\\
     & &(+6.3\%)&\textbf{(+7.1\%)}& &(+20.5\%)&\textbf{(+31.6\%)}& &(+1.9\%)&\textbf{(+3.7\%)}\\
     
    30\%&0.4876&0.4903&\textbf{0.4974}& 0.3728&0.3532&\textbf{0.3774}&0.6174&0.6304&\textbf{0.6334}\\
     & &(+0.5\%)&\textbf{(+2.0\%)}& &(-5.3\%)&\textbf{(+1.2\%)}& &(+2.1\%)&\textbf{(+2.6\%)}\\
     
    40\%&0.5090&0.5155&\textbf{0.5215}& 0.3993&0.3778&\textbf{0.4026}&0.6579&\textbf{0.6723}&0.6714\\
     & &(+1.3\%)&\textbf{(+2.5\%)}& &(-5.4\%)&\textbf{(+0.8\%)}& &\textbf{(+2.2\%)}&(+2.1\%)\\
     
    50\%&0.5233&0.5205&\textbf{0.5278}& 0.4290&0.3976&\textbf{0.4293}&0.6947&\textbf{0.7152}&0.7074\\
     & &(-0.5\%)&\textbf{(+0.8\%)}& &(-7.3\%)&\textbf{(+0.1\%)}& &\textbf{(+2.9\%)}&(+1.8\%)\\
     
    100\%&0.5592&0.5543&\textbf{0.5631}& 0.5017&0.4586&\textbf{0.5031}&0.8110&\textbf{0.8279}&0.8134\\
     & &(-0.9\%)&\textbf{(+0.7\%)}& &(-8.6\%)&\textbf{(+0.3\%)}& &\textbf{(+2.1\%)}&(+0.3\%)\\
   \bottomrule
\end{tabular}
\end{table*}

\begin{figure*}[h]
  \centering
  \includegraphics[width=\linewidth]{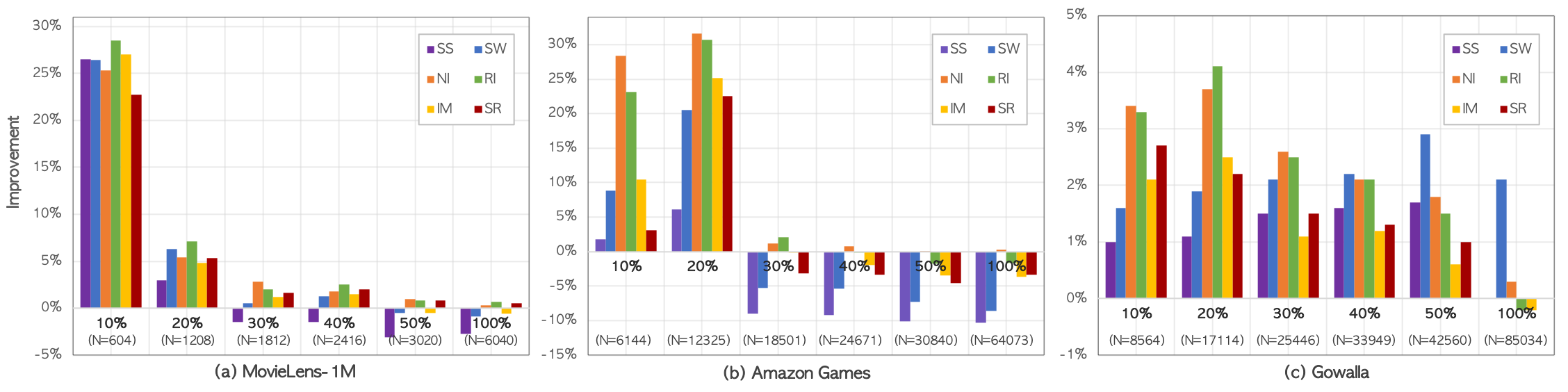}
  \caption{Performance Improvements based on the Baseline Model (i.e. SASRec) for Each Strategy (NDCG@10)}
  \label{fig:per_baseline}
\end{figure*}

Particularly, we aim to examine if data augmentation can help a sequential recommendation model generalize better when the amount of data is not sufficient, by assuming the low-resource situations where the training set size is intentionally restricted. Of course, it is easily inferred that there is no 'absolute' standard to determine if the amount of data is sufficient, which means it would be dependent on the characteristics of the dataset a system deals with. Regarding this, our extensive experiments show that data augmentation tends to be more effective on smaller datasets, but the criteria of low-resource situations to induce better generalization differ by the dataset.

{\itshape 4.5.2. Ablation Study: Effects of each strategy (RQ2).} Figure~\ref{fig:per_baseline} shows the improvements in performance for all experimental cases in this study. Most cases demonstrate performance improvement compared with the baseline model without augmentation. For total 71 cases\footnote{(3 datasets * 6 fractions * 4 strategies) - 1 = 71, excluding the case of SR for Gowalla full dataset due to its high computation cost} that our strategies are applied to, 78.9\verb|%| of them (i.e., 56 cases) are found to bring improvement. Specifically, it is obvious that data augmentation can be effective for smaller dataset but, for some strategies, can instead lead to worse performance when the size of the dataset is large enough.

Furthermore, it can be meaningful to explore why the effect is brought by each augmentation strategy. Among the direct manipulation strategies suggested, Noise/Redundancy Injection (NI/RI) shows the best performance improvements for most of datasets. We can infer that their excellence stems from grasping the skipping patterns of the short-term preferences in item sequences. According to the observations in previous studies, it is important to capture the union-level and skipping patterns of preferences as well as point-level ones (i.e., item-to-item transition) for enhancing performance in sequential recommendations\citep{Tang18,Yan19}. In other words, a model can yield better recommendations when it relaxes the structural constraint (i.e., one-directional chain-structure in ordered item sequences) than when it strictly maintains the constraint. Thus, we can conclude that these injection strategies lead to significantly better performance by capturing the skipping patterns in sequences through the injected data samples.

Besides, Synonym Replacement (SR) is shown to exhibit a substantial difference in improvement by the dataset, which may be brought by the difficulty of confirming how similar the computed synonyms are to the corresponding original, since the process of computing similar ones (i.e., synonyms) is not elaborately tuned in this experimental setting. Of course, in certain cases, synonyms may not fundamentally replace the original item. 

In case of prior works, Sliding Window (SW) shows more stable improvements than Subset Split (SS). We can infer that this is because SS, an item removal strategy, is bound to reduce the amount of available information. In the same regard, the relatively inferior performance of Item Masking (IM) strategy can be understood.

{\itshape 4.5.3. Ablation Study: Effects of data augmentation size (RQ3).} The experiments to see the variation of performance depending on the size of data augmentation are conducted only on the strategies of SW and NI to show the most outstanding improvement for each category, respectively. It turns out that the size of performance gain tends to be proportional to the size of augmentation for all datasets. Regardless of the selection of the dataset or augmentation strategies, it is hard to expect improvements in performance if the size of data augmentation is too small. We can conclude accordingly, for small data, noticeable improvements can be expected even with a small size of data augmentation while for sufficiently large data, a large size of data augmentation is required to generate meaningful improvement in performance. In other words, the effects of data augmentation become saturated as the amount of data becomes sufficiently large.

\begin{figure*}[h]
  \centering
  \includegraphics[width=\linewidth]{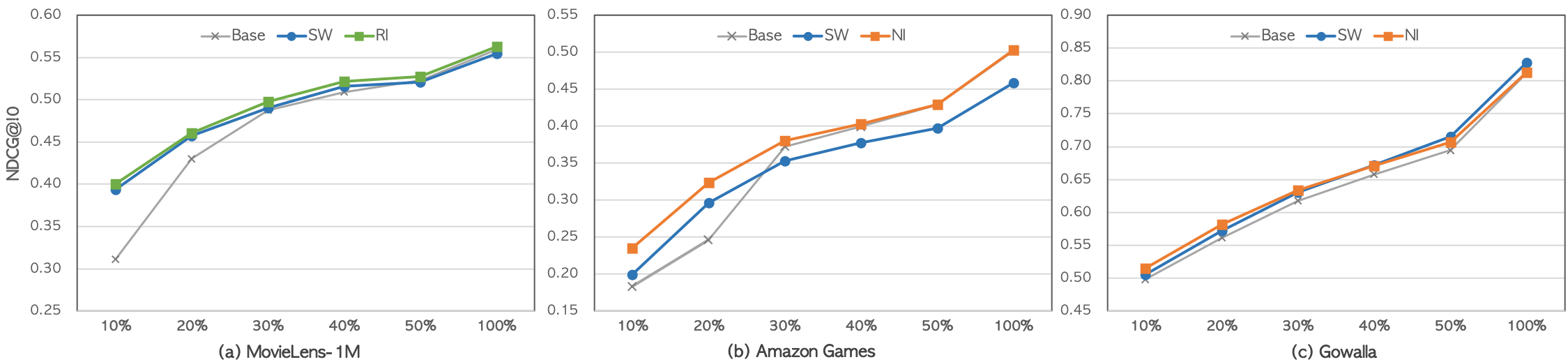}
  \caption{Saturation in Performance Improvement in line with the Increase in the Fraction of Sampled Dataset}
  \label{fig:saturation}
\end{figure*}

Table~\ref{tab:res_aug} summarizes the variation of the performance across the size of augmentation for all sub-datasets. Figure~\ref{fig:saturation} also shows the saturation of performance along with the increase in the training data size, when the size of augmentation is fixed $(N_{aug}=10)$. For MovieLens-1M and Amazon Games datasets, at the 10\verb|~|20\verb|%| of full datasets, the application of data augmentations results in significant performance improvements whereas the performances converge to that of the baseline models as the fraction increases to 100\verb|%|.

\begin{table*}
  \centering
  \caption{Amazon Games Performances across the Size of Data Augmentation $(N_{aug})$}
  \label{tab:res_aug}
  \begin{tabular}{c|c|cccc|cccc}
    \toprule
    Size & Base & \multicolumn{4}{c|}{Prior(SW)} & \multicolumn{4}{c}{Ours(NI)} \\
    \cline{2-10}
      & n=1 & n=2 & n=5 & n=10 & n=15 & n=2 & n=5 & n=10 & n=15 \\
    \midrule
    10\%&0.1830&0.1429&0.1790&0.1991&0.2076 &0.1532&0.2069&0.2350&0.2453\\
     & &(-21.9\%)&(-2.2\%)&(+8.8\%)&(+13.5\%)&(-16.4\%)&(+13.1\%)&(+28.4\%)&(+34.6\%)\\
     
    20\%&0.2462&0.1781&0.2577&0.2965&0.3168 &0.2233&0.2994&0.3081&0.3340\\
     & &(-27.7\%)&(+4.7\%)&(+20.5\%)&(+28.7\%)&(-9.3\%)&(+21.6\%)&(+25.2\%)&(+35.7\%)\\
     
    30\%&0.3728&0.2535&0.3357&0.3532&0.3665 &0.2919&0.3505&0.3774&0.3898\\
     & &(-32.0\%)&(-9.9\%)&(-9.0\%)&(-1.7\%)&(-21.7\%)&(-6.0\%)&(+1.2\%)&(+4.6\%)\\
     
    40\%&0.3993&0.2881&0.3630&0.3778&0.3868 &0.3247&0.3792&0.4026&0.4161\\
     & &(-27.9\%)&(-9.1\%)&(-5.4\%)&(-3.1\%)&(-18.7\%)&(-5.0\%)&(+0.8\%)&(+4.2\%)\\
     
    50\%&0.4290&0.3207&0.3869&0.3976&0.4050 &0.3522&0.4032&0.4293&0.4392\\
     & &(-25.2\%)&(-9.8\%)&(-7.3\%)&(-5.6\%)&(-17.9\%)&(-6.0\%)&(+0.1\%)&(+2.4\%)\\
     
    100\%&0.5017&0.3983&0.4445&0.4586&0.4666 &0.4250&0.4738&0.5031&0.5144\\
     & &(-20.6\%)&(-11.4\%)&(-8.6\%)&(-7.0\%)&(-15.3\%)&(-5.6\%)&(+0.3\%)&(+2.5\%)\\
   \bottomrule
\end{tabular}
\end{table*}

In general, it is relatively difficult to acquire additional data especially on the behavior logs of actual users for recommender systems. Thus, in the early stage of the system, the use of a relatively small dataset had intrinsic limitations as to achieving high recommendation performance, which inevitably caused low user satisfaction until substantial attainment of additional data. However, based on this result, we can strongly suggest that when a sequential recommender system does not have large enough data (i.e., a cold-start situation of the system), the application of data augmentation can more efficiently boost the performance than only focusing on the acquisition of additional data.

\section{Conclusion}

In this paper, we have shown that the application of data augmentation can contribute to boosting performance in DL-based sequential recommendations, especially in low-resource situations. Although the improvement may be marginal or negative at times, augmentation is found to help the model generalize better in most cases. Furthermore, we suggest a set of augmentation strategies that directly manipulate original sequences, distinctive from the prior approaches exploiting subsequences and prove that our strategies can result in better or competitive performance compared to the prior ones. Continued work on this topic could explore its more expanded application to other algorithms or network architectures such as CNN or VAE. We hope that our work can help manage cold-start situations for any early-stage real-world sequential recommender systems.

\bibliographystyle{unsrtnat}
\bibliography{1174_base}  

\appendix

\section{APPENDIX: Results based on HR@10 Metric}

\begin{table*}[h]
  \caption{Recommendation Performances with the Best Augmentation Strategies (HR@10)}
  \label{tab:res_main_hr}
  \begin{tabular}{c|ccc|ccc|ccc}
    \toprule
    Size& \multicolumn{3}{c|}{MovieLens-1M} & \multicolumn{3}{c|}{Amazon Games} & \multicolumn{3}{c}{Gowalla} \\
    \cline{2-10}
      & Base&Prior(SW)&Ours(RI)& Base&Prior(SW)&Ours(NI)& Base&Prior(SW)&Ours(NI) \\
    \midrule
    10\%& 0.5493&0.6473&\textbf{0.6507}&0.3184&0.3298&\textbf{0.3645}&0.5524&0.5667&\textbf{0.5720}\\
     & &(+17.8\%)&\textbf{(+18.5\%)}& &(+3.6\%)&\textbf{(+14.5\%)}& &(+2.6\%)&\textbf{(+3.6\%)}\\
     
    20\%& 0.6858&0.7050&\textbf{0.7066}&  0.4072&0.4599&\textbf{0.4868}& 0.6314&0.6465&\textbf{0.6555}\\
     & &(+2.8\%)&\textbf{(+3.0\%)}& &(+12.9\%)&\textbf{(+19.5\%)}& &(+2.4\%)&\textbf{(+3.8\%)}\\
     
    30\%& 0.7340&0.7336&\textbf{0.7418}& 0.5499&0.5374&\textbf{0.5558}& 0.6954&0.7112&\textbf{0.7156}\\
     & &(-0.1\%)&\textbf{(+1.1\%)}& &(-3.4\%)&\textbf{(+2.0\%)}& &(+2.3\%)&\textbf{(+2.9\%)}\\
     
    40\%& 0.7566&0.7611&\textbf{0.7635}& 0.5834&0.5685&\textbf{0.5895}& 0.7417&\textbf{0.7589}&0.7581\\
     & &(+0.6\%)&\textbf{(+0.9\%)}& &(-2.6\%)&\textbf{(+1.0\%)}& &\textbf{(+2.3\%)}&(+2.2\%)\\
     
    50\%& 0.7656&0.7656&\textbf{0.7728}& 0.6197&0.5970&\textbf{0.6212}& 0.7801&\textbf{0.8029}&0.7967\\
     & &( - )&\textbf{(+0.8\%)}& &(-3.7\%)&\textbf{(+0.2\%)}& &\textbf{(+2.9\%)}&(+2.1\%)\\
     
    100\%& 0.7929&0.7919&\textbf{0.7989}& 0.6968&0.6582&\textbf{0.6993}& 0.8968&\textbf{0.9139}&0.9028\\
     & &(-0.1\%)&\textbf{(+0.9\%)}& &(-5.5\%)&\textbf{(+0.4\%)}& &\textbf{(+1.9\%)}&(+0.7\%)\\
   \bottomrule
\end{tabular}
\end{table*}

\begin{figure*}[h]
  \centering
  \includegraphics[width=\linewidth]{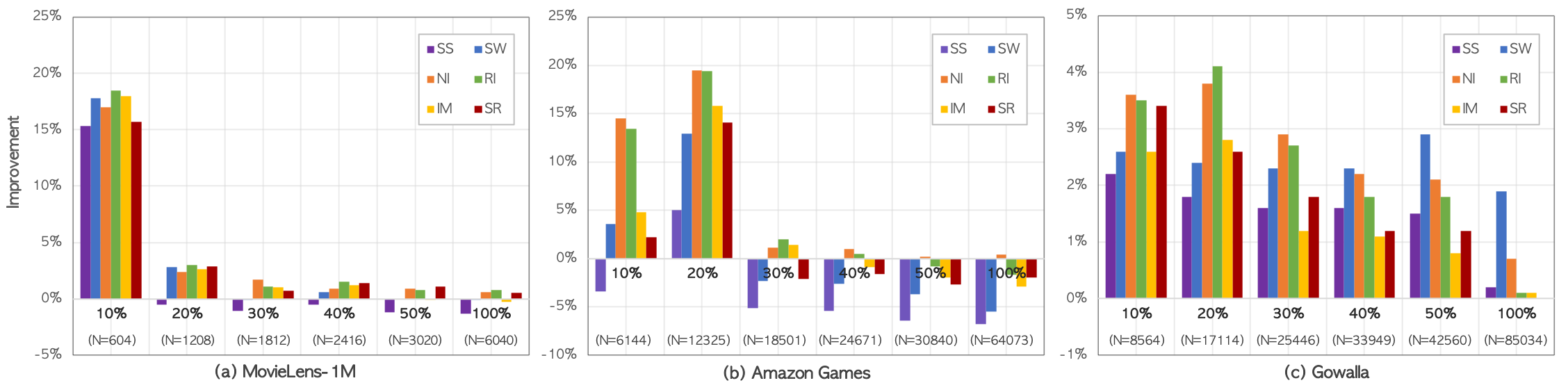}
  \caption{Performance Improvements based on the Baseline Model (i.e. SASRec) for Each Strategy (HR@10)}
  \label{fig:per_baseline_hr}
\end{figure*}

\end{document}